\documentstyle[12pt]{article}

\advance \textheight by \topskip
\def\ps#1{\raisebox{.2ex}{$\displaystyle
    \mathop{\psi}^{\scriptscriptstyle [#1]}$}{}}
\def\eps#1{\raisebox{.2ex}{$\displaystyle
    \mathop{\varepsilon}^{\scriptscriptstyle (#1)}$}{}}
\def\sig#1{\raisebox{.2ex}{$\displaystyle
    \mathop{\sigma}^{\scriptscriptstyle [#1]}$}{}}
\def\Lam#1{\raisebox{.2ex}{$\displaystyle
    \mathop{\Lambda}^{\scriptscriptstyle [#1]}$}{}}
\def\g#1{\raisebox{.2ex}{$\displaystyle
    \mathop{g}^{\scriptscriptstyle [#1]}$}{}}
\def\Ph#1{\raisebox{.2ex}{$\displaystyle
    \mathop{\Phi}^{\scriptscriptstyle [#1]}$}{}}
\def\u#1{\raisebox{.2ex}{$\displaystyle
    \mathop{u}^{\scriptscriptstyle [#1]}$}{}}
\def\A#1#2{{\mathop{A}\limits^{[#1]}_{[#2]}}{}}
\def\X#1#2{{\mathop{X}\limits^{[#1]}_{[#2]}}{}}
\def\Z#1#2{{\mathop{Z}\limits^{[#1]}_{[#2]}}{}}
\def\x#1{\raisebox{.2ex}{$\displaystyle
    \mathop{\xi}^{\scriptscriptstyle [#1]}$}{}}

\makeatletter
\@addtoreset{equation}{section}

\makeatother

\begin{document}

\begin{titlepage}
\hbox to \hsize{\hfil hep-th/9407156}
\hbox to \hsize{\hfil INR 860/94}
\hbox to \hsize{\hfil July, 1994}
\vfill
\large \bf
\begin{center}
CONSTRAINT ALGEBRAS \\
IN GAUGE INVARIANT SYSTEMS
\end{center}
\vskip1cm
\normalsize
\begin{center}
{\bf Kh. S. Nirov\footnote{E--mail: nirov@amber.inr.free.net}}\\
{\small Institute for Nuclear Research of the Russian Academy of
Sciences} \\
{\small 60th October Anniversary prospect 7a, 117312 Moscow, Russia}
\end{center}
\vskip2cm
\begin{abstract}
\noindent
The Hamiltonian description for a wide class of mechanical systems,
having local symmetry trans\-for\-ma\-tions depending
on time derivatives of the gauge parameters of arbitrary order,
is constructed.  The Poisson
brackets of the Hamiltonian and constraints with each other and with
arbitrary function are explicitly obtained. The constraint algebra is
proved to be the first class.
\end{abstract}
\vfill
\end{titlepage}

\section{Introduction}

Gauge invariance gives rise to constraints of both Lagrangian and
Hamiltonian formalisms (see \cite{GT,Nest,RS,PR,LL,NPR,P} and references
therein). Lagrangian constraints appear as a consequence of a functional
dependence of the Lagrange equations. In general, there exist
gauge invariant systems having no Lagrangian constraints. These
systems correspond to the gauge transformations of the form
\begin{equation}
\delta_\varepsilon q^r = \varepsilon^\alpha \psi^r_\alpha(q,\dot q) .
\label{1.0}
\end{equation}
But for all known systems of such type it is possible to find an
equivalent gauge invariant system with Lagrangian constraints.
Such a situation is observed e.~g. for relativistic particles and strings.

To construct the Hamiltonian description
of the system, it is necessary to use the Legendre transformation, but for
the case of gauge invariant systems it turns out to be a singular mapping.
This fact leads to the constraints in Hamiltonian formalism. Thus,
since the origins of Lagrangian and Hamiltonian constraints
seem to be different, the questions of the correspondence between
Lagrangian and Hamiltonian descriptions of gauge invariant systems,
the connection between Lagrangian and Hamiltonian constraints, are
risen (see \cite{Isp1,Isp2} and refs. therein).

The structure of the theory (i.e. the Noether identities, gauge algebras,
hierarchy of constraints, etc.) is determined by the type of gauge
transformations. In particular, the systems of Yang--Mills type \cite{NR1}
are characterized by the gauge symmetry transformations of trajectory
\begin{equation}
\delta_\varepsilon q^r(t) = \varepsilon^\alpha(t) \xi^r_\alpha(q(t)) +
\dot\varepsilon^\alpha(t) \psi^r_\alpha(q(t)), \label{1.1}
\end{equation}
whereas the systems describing theories of gravity, strings,
relativistic particles have the local symmetry under transformations
of the form \cite{PR,NPR,NR2,NR3}
\begin{equation}
\delta_\varepsilon q^r(t) = \varepsilon^\alpha(t)\left(\x{0}(q(t)) +
\dot{q}^s(t)\x{1}^r_{\alpha s}(q(t))\right) +
\dot\varepsilon^\alpha(t)\psi^r_\alpha(q(t)) . \label{1.2}
\end{equation}
We see that various dependences on the velocity phase space coordinates
in the gauge transformations lead to essentially different physical theories.

All of the most interesting from the physical point of view systems,
as we know, correspond to the gauge transformations depending only on
(up to) first order time derivative of infinitesimal gauge parameters
$\varepsilon^\alpha(t)$. Note also that the most general form of the gauge
symmetry transformations for the quadratic systems are given by (\ref{1.2})
\cite{NR2}.
Now we shall generalize consideration of Refs.\cite{PR,NPR} to the class
of the gauge symmetry transformations with higher (arbitrary) order time
derivatives of infinitesimal gauge parameters.

The paper is organized as follows. In Section 1 the Lagrangian formalism for
systems, invariant under gauge transformations of a general form, but
depending only on the velocity phase space coordinates, is presented.
Section 2 is devoted to the Hamiltonian description of such systems;
using the Noether identities and the gauge algebra relations obtained
in Section 1, we get the Poisson brackets of the Hamiltonian and
constraints on the primary constraint surface. The correspondence between
Lagrangian and Hamiltonian formalisims is clarified.
In Section 3, using the notion of the standard extension \cite{PR,NPR,NR3},
we find the explicit form of the constraint algebras in the total phase space.
Besides, we calculate the Poisson brackets of the Hamiltonian and constraints
with an arbitrary function on the phase space; these expressions
may be useful for some applications.

Here we consider only bosonic mechanical systems. Note that one
can easily generalize the results of the paper to the case of
mechanical systems, described by both even and odd variables
\cite{PRR,DeW}.

The summation over repeated indexes is assumed.

\section{Gauge invariance in Lagrangian  mechanics}

Let us consider mechanical system given by Lagrangian $L(q,\dot q)$
in the $2R$--dimensional velocity phase space \cite{Arn}
with coordinates $q^r$, $\dot q^r$, $r = 1,\ldots,R$.
Hereafter, $q^r$ and $\dot q^r$ are
the generalized coordinates and the generalized velocities of the system,
respectively. It is convenient to present the Lagrange equations as follows:
\begin{equation}
L_r(q,\dot q,\ddot q) \equiv W_{rs}(q,\dot q)\ddot q^s - R_r(q,\dot q) = 0,
\label{2.1}
\end{equation}
where
\begin{eqnarray}
R_r(q,\dot q) &=& \frac{\partial L(q,\dot q)}{\partial q^r} -
\dot q^s\frac{\partial^2 L(q,\dot q)}{\partial q^s\partial \dot q^r},
\label{2.2} \\
W_{rs}(q,\dot q) &=& \frac{\partial^2 L(q,\dot q)}{\partial \dot q^r \dot q^s}.
\label{2.3}
\end{eqnarray}
The matrix $W_{rs}$ is called the Hessian of the system.

Assume that the system has a gauge symmetry under infinitesimal
trajectory transformations of the form
\begin{equation}
\delta_\varepsilon q^r = \sum_{k=0}^N
{\eps{k}^\alpha \ps{N-k}^r_\alpha(q,\dot q)},
\qquad \alpha = 1,\ldots,A,
\label{2.4}
\end{equation}
where $\varepsilon^\alpha$ are arbitrary infinitesimal functions of time:
\begin{equation}
\delta_\varepsilon L = \frac{d}{dt} \Sigma_\varepsilon .
\label{2.5}
\end{equation}
In this paper we use the notations: integers within
parentheses over characters display an order of time derivative of
corresponding functions, and all the integers within square brackets
(both subscripts and superscripts of characters) just mark the functions,
simply giving them numbering.
Hence, the integer $N$ is the maximal order of time derivatives of the gauge
parameters $\varepsilon^\alpha(t)$ for the gauge invariant system we consider.
The case of $N = 1$, which is the most interesting from the physical point of
view, was considered in Refs.\cite{PR,NPR}. Now we shall treat the case of
arbitrary $N > 1$.

{}From the symmetry equations (\ref{2.4}),(\ref{2.5}) we get the Noether
identities
\begin{equation}
\sum_{k=0}^N {(-1)^k \frac{d^k}{dt^k} \left(\ps{N-k}^r_\alpha L_r\right)} = 0 .
\label{2.6}
\end{equation}
Besides, for $\Sigma_\varepsilon$ we have
\begin{equation}
\Sigma_\varepsilon = \sum_{k=0}^N {\eps{k}^\alpha \sig{N-k}_\alpha} ,
\label{2.7}
\end{equation}
where
\begin{equation}
\sig{k}_\alpha = \ps{k}^r_\alpha \frac{\partial L}{\partial \dot q^r} -
\sum_{l=0}^k {(-1)^l \frac{d^l}{dt^l} \left(\ps{k-1-l}^r_\alpha L_r\right)} .
\label{2.8}
\end{equation}
One can rewrite the Noether identities (\ref{2.6}) in the following equivalent
form
\begin{eqnarray}
\Lam{k+1}_\alpha &=& \ps{k}^r_\alpha R_r - \dot q^s
\frac{\partial \Lam{k}_\alpha}{\partial q^s} ,
\label{2.9} \\
\ps{k}^r_\alpha W_{rs} &=&
- \frac{\partial \Lam{k}_\alpha}{\partial \dot q^s} ,
\label{2.10}
\end{eqnarray}
where $k = 0,1,\ldots,N$ and $\Lam{0}_\alpha = \Lam{N+1}_\alpha \equiv 0$.
Using the Noether identities in the form of (\ref{2.9})--(\ref{2.10}), we get
more convenient form of $\Sigma_\varepsilon$, namely
\begin{equation}
\Sigma_\varepsilon = \delta_\varepsilon q^r
\frac{\partial L}{\partial \dot q^r}
+ \sum_{k=0}^{N-1} {\eps{k}^\alpha \Lam{N-k}_\alpha} .
\label{2.11}
\end{equation}

It follows from Eq.(\ref{2.10}) that the Hessian of the system has $A$
null vectors $\ps{0}^r_\alpha$, $\alpha = 1,\ldots,A$. Suppose that the
vectors $\ps{0}^r_\alpha$ are linearly independent, and any null vector of
the matrix $W_{rs}$ is a linear combination of the vectors $\ps{0}^r_\alpha$.
Hence, we have
\begin{equation}
{\rm rank}\; W_{rs}(q,\dot q) = R - A , \qquad
{\rm rank}\; \ps{0}^r_\alpha(q,\dot q) = A  \label{2.12}
\end{equation}
for any values of $q^r$ and $\dot q^r$. We consider the systems for which
any choice of arbitrary functions $\varepsilon^\alpha(t)$ and any trajectory
gives the gauge transformations (\ref{2.4}) to be nontrivial. One can show
that this condition is equivalent to the linear independence of the set
formed by the vectors $\ps{k}^r_\alpha$, $k = 0,1,\ldots,N$.

The rank of the Hessian is less than dimension of the configuration space of
the system, hence, the Cauchy problem for the Lagrange equations (\ref{2.1})
has no unique solution and there are intersecting trajectories in the
system \cite{GT}. In other words, using the Lagrange equations, we can
express only $R - A$ accelerations $\ddot q^r$ through the generalized
coordinates and the generalized velocities. The remaining equations have the
form
\begin{equation}
\ps{0}^r_\alpha R_r = 0 , \label{2.13}
\end{equation}
that follows directly from Eqs.(\ref{2.1}),(\ref{2.10}). Such relations
restrict the possible values of $q^r$ and $\dot q^r$ and are called the
primary Lagrangian constraints.

Using the Noether identities we get from the stability condition for the
primary Lagrangian constraints $\Lam{1}_\alpha = \ps{0}^r_\alpha R_r$ other
Lagrangian constraints of the system, which are the relations
$\Lam{k}_\alpha = 0$, $k = 2,\ldots,N$. The Lagrangian constraints of
$(k+1)$-th stage $\Lam{k+1}_\alpha$ appear as a consequence of stability
(with respect to time evolution) of the preceding Lagrangian constraints of
$k$-th stage $\Lam{k}_\alpha$.

Suppose now that gauge transformations (\ref{2.4}) form a closed gauge
algebra. So, for any two sets of infinitesimal functions
$\varepsilon^\alpha_1(t)$ and $\varepsilon^\alpha_2(t)$ we have the commutator
of corresponding gauge transformations of type (\ref{2.4}) to be of the same
type
\begin{equation}
\left[\delta_{\varepsilon_1} \,\
\delta_{\varepsilon_2}\right] q^r = \delta_\varepsilon q^r .
\label{2.14}
\end{equation}
In this, $\varepsilon^\alpha$ are, in general, some functions of
$\varepsilon^\alpha_1$, $\varepsilon^\alpha_2$ and the trajectory of the
system.

Using (\ref{2.4}) and taking into account the linear independence of the
vectors $\ps{k}^r_\alpha$, $k = 0,1,\ldots,N$, we obtain from Eq.(\ref{2.14})
the relations
\begin{eqnarray}
 \ps{N-m+n}^s_\alpha \frac{\partial \ps{N-n}^r_\beta}{\partial q^s}
&+& \left( \ps{N-m+n+1}^s_\alpha + \dot{\ps{N-m+n}^s_\alpha} \right)
\frac{\partial \ps{N-n}^r_\beta}{\partial \dot q^s}  \nonumber \\
- \ps{N-n}^s_\beta \frac{\partial \ps{N-m+n}^r_\alpha}{\partial q^s}
&-& \left( \ps{N-n+1}^s_\beta + \dot{\ps{N-n}^s_\beta} \right)
\frac{\partial \ps{N-m+n}^r_\alpha}{\partial \dot q^s}  \nonumber \\
= \sum_{i=0}^N \sum_{j=0}^i
{\left( \begin{array}{c}
         i \\ j
 \end{array} \right)} \A{n-j}{m-i}^\gamma_{\alpha \beta}\,\ps{N-i}^r_\gamma
&+& \left[ \dot\A{n}{m}^\gamma_{\alpha \beta} \ps{1}^r_\gamma + \left(
\ddot\A{n}{m}^\gamma_{\alpha \beta} + 2 \dot\A{n}{m-1}^\gamma_{\alpha
\beta} + 2 \dot\A{n-1}{m-1}^\gamma_{\alpha \beta} \right)
\ps{0}^r_\gamma \right]  ,
\label{2.15}
\end{eqnarray}
where $n = 0,1,\ldots,N+1$; $m = 0,1,\ldots,2N+1$.
Here $\A{l}{k}^\gamma_{\alpha \beta}$ are some functions of the
generalized coordinates $q^r$, called the structure functions of the
gauge algebras (remember that for the case of $N = 1$ \cite{PR,NPR}
the structure functions of the corresponding gauge algebras, in
general, depend on both generalized coordinates and generalized
velocities).

These functions satisfy the symmetry equations
\begin{equation}
\A{l}{k}^\gamma_{\alpha \beta} = - \A{k-l}{k}^\gamma_{\beta \alpha} ,
\qquad l \le k \le l + 1 , \label{2.16}
\end{equation}
and relate the infinitesimal parameters of gauge transformations in
(\ref{2.14}) as follows
\begin{equation}
\varepsilon^\gamma = \sum_{k=0}^{N+1} \sum_{l=0}^k
\eps{k-l}^\alpha_1\,\,\eps{l}^\beta_2\,\,\A{l}{k}^\gamma_{\alpha \beta} .
\label{2.17}
\end{equation}

However, we get from (\ref{2.16}) that only the structure functions
$\A{0}{0}^\gamma_{\alpha \beta}$, $\A{0}{1}^\gamma_{\alpha \beta}$,
$\A{1}{1}^\gamma_{\alpha \beta}$, $\A{1}{2}^\gamma_{\alpha \beta}$
are nonzero.  Besides, as follows from Eq.(\ref{2.15}), for the
cases of $N > 2$  all the structure functions are turned out to be
constant, thus the terms within the square brackets in the r.~h.~s. of
(\ref{2.15}) are equal to zero. These properties of the gauge algebras
essentially simplify the further analysis.

Note also that Eq.(\ref{2.15}) contains the relations
\begin{equation}
\ps{0}^s_\alpha \frac{\partial \ps{n}^r_\beta}{\partial \dot q^s}
= \A{N-n}{N-n+1}^\gamma_{\alpha \beta}\,\,\ps{0}^r_\gamma ,
\qquad n = 0,1,\ldots,N .
\label{2.18}
\end{equation}
We see that the r.~h.~s. of Eq.(\ref{2.18}) for values of $n < N - 1$
is zero because of the above properties of the structure functions.

Thus, we have obtained in this Section the Noether identities, the gauge
algebra relations and the hierarchy of the Lagrangian constraints
\begin{equation}
\Lam{k}_\alpha = 0 , \qquad k = 1,\ldots,N .  \label{2.19}
\end{equation}
These expressions will be used in the next Section to construct Hamiltonian
description for the systems under consideration.

\section{Hamiltonian description of gauge invariant systems}

To write down Hamiltonian description of the systems under
consideration, one needs the algebras of constraints and the
Hamiltonian. It forces us to find out the correspondence between
Lagrangian and Hamiltonian constraints in the spirit of
Refs.\cite{PR,NPR}.

To realize a Hamiltonian formalism, let us introduce $2R$--dimensional
phase space described by canonical pairs of the generalized
coordinates $q^r$ and generalized momenta $p_r$, and define a mapping
of the velocity phase space to the (canonical) phase space with the
help of usual relation
\begin{equation} p_r(q,\dot q) = \frac{\partial
L(q,\dot q)}{\partial \dot q^r} . \label{3.1}
\end{equation}
As it was established in the previous Section, the
Hessian of the system
$W_{rs}(q,\dot q) = {\partial p_r(q,\dot q)}/{\partial \dot q^s}$
is singular, hence the mapping given by
Eq.(\ref{3.1}) has no inverse. One can show \cite{RS} that under this
mapping an inverse image of a point of the phase space is either
empty or consists of one or several $A$--dimensional surfaces having
the parametric representation of the form \begin{eqnarray}
q^r(\tau) &=& q^r , \label{3.2} \\ \dot q^r(\tau) &=& \dot q^r +
\tau^\alpha \ps{0}^r_\alpha(q,\dot q) .  \label{3.3}
\end{eqnarray} Taking this fact into account and disregarding the
degenerative cases, we see that the image of the velocity phase space
under the mapping (\ref{3.1}) is a $(2R - A)$--dimensional surface in
the phase space, which may be given with the help of $A$ functionally
independent functions as follows \begin{equation}
\Ph{0}_\alpha (q,p) = 0 , \qquad \alpha = 1,\ldots,A . \label{3.4}
\end{equation}
Hence, we have introduced irreducible set of the so--called primary
constraints $\Ph{0}_\alpha$ \cite{Dir} and defined the primary constraint
surface by Eq.(\ref{3.4}).

For any function $F(q,p)$, defined on the phase space, we can
introduce the corresponding function $f(q,\dot q)$ on the velocity
phase space by the relation
\begin{equation}
f(q,\dot q) = F(q,p(q,\dot q)) . \label{3.5}
\end{equation}
It follows from the relation
\begin{equation}
p_r(q(\tau),\dot q(\tau)) = p_r(q,\dot q) ,
\label{3.6}
\end{equation}
where $\tau^\alpha$ parametrize the
surfaces (\ref{3.2}), (\ref{3.3}), that the function $f(q,\dot q)$ is
constant on such surfaces. This fact implies differential expressions
of the form
\begin{equation}
\ps{0}^r_\alpha \frac{\partial f}{\partial \dot q^r} = 0 ,
\qquad \alpha = 1,\ldots,A . \label{3.7}
\end{equation}
However, given a function $f(q,\dot q)$ on the
velocity phase space, it is not always possible to define a function
$F(q,p)$ on the phase space, connected with $f$ by the relation
\begin{equation}
F(q,p(q,\dot q)) = f(q,\dot q) . \label{3.8}
\end{equation}
The necessary conditions for the existence of such a
function are (\ref{3.7}).  These conditions become sufficient if each
point of the primary constraint surface is the image of only one
surface of form (\ref{3.2}), (\ref{3.3}).  We suppose that it is
really so.

Thus, in the case under consideration for any function $f(q,\dot q)$,
satisfying the conditions (\ref{3.7}), we can find a function $F(q,p)$,
such that Eq.(\ref{3.8}) is fulfilled. When the functions $F$ and $f$ are
connected by the relation (\ref{3.8}), we shall use the notations of
Refs.\cite{PR,NPR} and write
\begin{equation}
F \doteq f . \label{3.9}
\end{equation}
We shall also call such a function $f$ projectable to the primary constraint
surface, or simply projectable.

Note that
\begin{equation}
\Ph{0}_\alpha \doteq 0 , \label{3.10}
\end{equation}
thus, if a function $F_0$ satisfies (\ref{3.9}), then any function $F$
of the form
\begin{equation}
F = F_0 + F^\alpha \Ph{0}_\alpha , \label{3.11}
\end{equation}
where $F^\alpha$ are arbitrary functions, satisfies this equality as
well.  Treating (\ref{3.9}) as an equation for $F$, and $F_0$ as a
partial solution of this equation, one can show that (\ref{3.11})
gives the general solution of this equation. In other words, the
relation (\ref{3.9}) defines the function $F$ only on the primary
constraint surface, and it may be extended from this surface to the
total phase space arbitrarily, but according to (\ref{3.11}) various
extensions will differ from each other in a linear combination of the
primary constraints \cite{RS}.

Now let us define the Hamiltonian of the system. To this end,
introduce the energy function $E(q,\dot q)$
on the velocity phase space  by the relation
\begin{equation}
E = \dot q^r \frac{\partial
L}{\partial \dot q^r} - L . \label{3.12}
\end{equation}
The energy function is projectable to the primary constraint surface,
hence we can find the corresponding function $H(q,p)$ on the phase
space, such that \begin{equation} H \doteq E , \label{3.13}
\end{equation} and treat this function as the Hamiltonian of the
system. According to the above reasonings, Eq.(\ref{3.13}) determines
the Hamiltonian only at the points of the primary constraint surface.
The function $H(q,p)$ as function on the total phase space may be
obtained by arbitrary extensions from this surface.

Recall now that the Lagrangian constraints $\Lam{k}_\alpha(q,\dot q)$,
$k = 1,\ldots,N$, take constant values on the surfaces of the form (\ref{3.2}),
(\ref{3.3}), i.~e. they satisfy the conditions (\ref{3.7}). It follows
directly from the Noether identities (\ref{2.10}). Hence, one can find on the
phase space the functions $\Ph{k}_\alpha(q,p)$, such that
\begin{equation}
\Ph{k}_\alpha \doteq \Lam{k}_\alpha , \qquad k = 1,\ldots,N . \label{3.14}
\end{equation}

Let us compute the Poisson brackets of the Hamiltonian $H$, primary
constraints $\Ph{0}_\alpha$ and functions $\Ph{k}_\alpha$, $k = 1,\ldots,N$,
corresponding to the Lagrangian constraints. To this end, we shall obtain
the partial derivatives of all these functions over the canonical coordinates
and momenta.

One can write the explicit form of Eq.(\ref{3.10})
\begin{equation}
\Ph{0}_\alpha (q,p(q,\dot q)) = 0 . \label{3.15}
\end{equation}
Differentiating these relations over $\dot q^r$ and $q^r$, respectively, we
get the following expressions
\begin{eqnarray}
\frac{\partial \Ph{0}_\alpha}{\partial p_r} &\doteq&
- \u{0}^\beta_\alpha \ps{0}^r_\beta , \label{3.16} \\
\frac{\partial \Ph{0}_\alpha}{\partial q^r} &\doteq&
 \u{0}^\beta_\alpha \ps{0}^s_\beta
 \frac{\partial^2 L}{\partial \dot q^s \partial q^r}, \label{3.17}
\end{eqnarray}
where the matrix $\u{0}^\beta_\alpha(q,\dot q)$ has to be invertible
since the primary constraints are functionally independent.

Taking into account (\ref{2.18}) we obtain that the matrix elements of
$\u{0}^\beta_\alpha(q,\dot q)$ are projectable to the primary constraint
surface:
\begin{equation}
\ps{0}^r_\alpha \frac{\partial \u{0}^\gamma_\beta}{\partial \dot q^r} = 0 .
\label{3.18}
\end{equation}
For any choice of the primary constraints, which define one and the same
primary constraint surface, we shall get different matrices
$\u{0}^\beta_\alpha$, satisfying Eq.(\ref{3.18}).

Differentiating the energy function $E(q,\dot q)$ over $q^r$ and $\dot q^r$
we get
\begin{eqnarray}
\frac{\partial E}{\partial \dot q^r} &=& \dot q^s W_{sr} ,
\label{3.19} \\
\frac{\partial E}{\partial q^r} &=& \dot q^s
\frac{\partial^2 L}{\partial \dot q^s \partial q^r} -
\frac{\partial L}{\partial q^r} .
\label{3.20}
\end{eqnarray}
Taking into account (\ref{3.13}), we obtain for the partial derivatives of the
Hamiltonian the expressions
\begin{eqnarray}
\frac{\partial H}{\partial p_r} &\doteq&
\dot q^r - \mu^\alpha \ps{0}^r_\alpha ,
\label{3.21} \\
\frac{\partial H}{\partial q^r} &\doteq& - \frac{\partial L}{\partial q^r} +
\mu^\alpha \ps{0}^s_\alpha
\frac{\partial^2 L}{\partial \dot q^s \partial q^r} ,
\label{3.22}
\end{eqnarray}
where the functions $\mu^\alpha(q,\dot q)$  satisfy the equalities
\begin{equation}
\ps{0}^r_\alpha \frac{\partial \mu^\beta}{\partial \dot q^r} =
\delta^\beta_\alpha . \label{3.23}
\end{equation}

Using the Noether identities one can obtain the partial derivatives of the
functions $\Ph{k}_\alpha$ from Eq.(\ref{3.14}). We have for any
$k = 1,\ldots,N$
\begin{eqnarray}
\frac{\partial \Ph{k}_\alpha}{\partial p_r} &\doteq& - \ps{k}^r_\alpha +
\u{k}^\beta_\alpha \ps{0}^r_\beta , \label{3.24} \\
\frac{\partial \Ph{k}_\alpha}{\partial q^r} &\doteq&
\frac{\partial \Lam{k}_\alpha}{\partial q^r} + \left( \ps{k}^s_\alpha -
\u{k}^\beta_\alpha \ps{0}^s_\beta \right)
\frac{\partial^2 L}{\partial \dot q^s \partial q^r} , \label{3.25}
\end{eqnarray}
where the functions $\u{k}^\beta_\alpha(q,\dot q)$ fulfil the relations
\begin{equation}
\ps{0}^r_\alpha \frac{\partial \u{k}^\gamma_\beta}{\partial \dot q^r} =
\A{N-k}{N-k+1}^\gamma_{\alpha \beta} , \label{3.26}
\end{equation}
following from Eq.(\ref{2.18}).

Note that the arbitrariness in the choice of functions $\mu^\alpha$ and
$\u{k}^\beta_\alpha$ is a consequence of the ambiguity of the extension
of the Hamiltonian and functions, corresponding to the Lagrangian constraints,
from the primary constraint surface. But for all the possible extensions
these functions must satisfy Eqs.(\ref{3.23}) and (\ref{3.26}),
respectively.

Using the relations of the gauge algebra (\ref{2.15})--(\ref{2.18}) and
the Noether identities (\ref{2.9}), (\ref{2.10}), we get from
Eqs.(\ref{3.16}), (\ref{3.17}), (\ref{3.21}), (\ref{3.22}), (\ref{3.24}),
(\ref{3.25}) the following expressions for the Poisson brackets
\begin{eqnarray}
\{ H\,,\,\Ph{0}_\alpha \} &\doteq& \u{0}^\beta_\alpha\;\Lam{1}_\beta ,
\label{3.27} \\
\{ H\,,\,\Ph{k}_\alpha \} &\doteq& \Lam{k+1}_\alpha -
\left( \u{k}^\beta_\alpha + \mu^\delta
\A{1}{N-k+1}^\beta_{\alpha \delta} \right)
\Lam{1}_\beta . \label{3.28}
\end{eqnarray}
We see that the Poisson brackets of the Hamiltonian with the functions
$\Ph{k}_\alpha, k = 0,1,\ldots,N$, give rise to the Lagrangian constraints
of the next order $\Lam{k+1}_\alpha$. This fact is in accordance with the
Lagrangian approach, given in the previous Section.

It is not also difficult to obtain the expressions for the Poisson
brackets of the primary constraints $\Ph{0}_\alpha$ with each other
and with the functions $\Ph{k}_\alpha$.
We have for any $k = 1,\ldots,N$ that
\begin{eqnarray}
\{ \Ph{0}_\alpha\,,\,\Ph{0}_\beta \} &\doteq& 0 , \label{3.29} \\
\{ \Ph{k}_\alpha\,,\,\Ph{0}_\beta \} &\doteq& \u{0}^\delta_\beta\,
\A{1}{N-k+1}^\gamma_{\alpha \delta}\,\Lam{1}_\gamma , \label{3.30}
\end{eqnarray}
where we again made use of the Noether identities, the gauge algebra
relations and the expressions for the partial derivatives (\ref{3.16}),
(\ref{3.17}) and (\ref{3.24}), (\ref{3.25}).

Taking into account Eqs.(\ref{3.27})--(\ref{3.30}) we see that
the functions $\Ph{k}_\alpha$, introduced by Eq.(\ref{3.14}), are nothing
but the secondary Hamiltonian constraints of $k$-th stage \cite{Dir,GT}.

The most difficult calculational problem here is to obtain the Poisson
brackets of the functions $\Ph{k}_\alpha$, $k = 1,\ldots,N$, with each other.
Using the above results we get from Eqs.(\ref{3.24}), (\ref{3.25}), that
\begin{equation}
\{ \Ph{k}_\alpha\,,\,\Ph{l}_\beta \} \doteq \left( \u{k}^\delta_\alpha\,
\A{1}{N-l+1}^\gamma_{\beta \delta} - \u{l}^\delta_\beta\,
\A{1}{N-k+1}^\gamma_{\alpha \delta} \right) \Lam{1}_\gamma +
\X{k}{l}_{\alpha \beta}  \label{3.31}
\end{equation}
for any $k,\,l = 1,\ldots,N$, where we have introduced the notation
\begin{equation}
\X{k}{l}_{\alpha \beta} =
\ps{k}^r_\alpha \frac{\partial \Lam{l}_\beta}{\partial q^r}
- \ps{l}^r_\beta \frac{\partial \Lam{k}_\alpha}{\partial q^r} -
\ps{k}^s_\alpha \ps{l}^r_\beta \left( \frac{\partial R_r}{\partial \dot q^s}
+ \dot q^t \frac{\partial W_{rs}}{\partial q^t} \right) . \label{3.32}
\end{equation}

Recall that for $N > 2$ from the gauge algebra it follows that
the structure functions are constant on the velocity
phase space. Now, considering the case of $N > 2$ we get, after tiresome
calculations, that the functions $\X{k}{l}_{\alpha \beta}$,
$k,l = 1,\ldots,N$, satisfy the following recursive relations
\begin{eqnarray}
&&\X{k}{l}_{\alpha \beta} +
\frac{1}{2} \X{k+1}{l-1}_{\alpha \beta} + \frac{1}{2} \X{k-1}{l+1}_{\alpha
\beta} + \frac{1}{2} \dot q^r \frac{\partial}{\partial q^r}
\X{k-1}{l}_{\alpha \beta} + \frac{1}{2} \dot q^r \frac{\partial}{\partial q^r}
\X{k}{l-1}_{\alpha \beta} \nonumber \\ &=& \frac{1}{2} \sum_{i=0}^N
\sum_{j=0}^i {\left( \begin{array}{c} i \\ j \end{array} \right)}
\left( \A{N-l-j}{2N+1-k-l-i}^\gamma_{\alpha \beta} +
       \A{N+1-l-j}{2N+1-k-l-i}^\gamma_{\alpha \beta} \right)
\ps{N-i}^r_\gamma R_r .
\label{3.33}
\end{eqnarray}
Using the relations (\ref{2.9}) and the properties (\ref{2.16}) of the
structure functions, we get the recursive relations (\ref{3.33}) to be
equivalent to the recursive equations of a more simple form
\begin{equation}
\X{k}{l}_{\alpha \beta} + \frac{1}{2} \X{k+1}{l-1}_{\alpha \beta} +
\frac{1}{2} \X{k-1}{l+1}_{\alpha \beta} =  \Z{k}{l}_{\alpha \beta} ,
\label{3.34}
\end{equation}
where
\begin{equation}
\Z{k}{l}_{\alpha \beta} = \frac{1}{2} \sum_{i=0}^2 \sum_{j=0}^i {\left(
                          \begin{array}{c}
                          2N+2-k-l-i \\ N+1-l-j
                          \end{array} \right)}
\A{j}{i}^\gamma_{\alpha \beta}\,\Lam{k+l-N+i}_\gamma . \label{3.35}
\end{equation}
We obtain that the solution to the recursive relations (\ref{3.34}) is given
by the expression
\begin{eqnarray}
&&(k+l) \X{k}{l}_{\alpha \beta} = (-1)^kl\,\X{0}{k+l}_{\alpha \beta} +
(-1)^lk\,\X{k+l}{0}_{\alpha \beta} + 2kl\,\Z{k}{l}_{\alpha \beta}
\nonumber \\
&+& 2l \sum_{m=1}^{k-1} {(-1)^m (k-m)\,
\Z{k-m}{l+m}_{\alpha \beta}} + 2k \sum_{m=1}^{l-1} {(-1)^m (l-m)\,
\Z{k+m}{l-m}_{\alpha \beta}} , \label{3.36}
\end{eqnarray}
where we have used the notations
\begin{equation}
\X{0}{k+l}_{\alpha \beta} = \A{N-k-l}{N-k-l+1}^\gamma_{\alpha \beta}\,
\Lam{1}_\gamma ,
\qquad \X{k+l}{0}_{\alpha \beta} = \A{1}{N-k-l+1}^\gamma_{\alpha \beta}\,
\Lam{1}_\gamma .
\label{3.37}
\end{equation}
Using the explicit form of the functions $\Z{k}{l}_{\alpha \beta}$, given by
Eq.(\ref{3.35}), we finally obtain
\begin{equation}
\X{k}{l}_{\alpha \beta} = \sum_{i=0}^2 \sum_{j=0,1} {\left(
                                  \begin{array}{c}
                                    2N-k-l-i \\ N-l-j
                                  \end{array} \right)}
\A{j}{i}^\gamma_{\alpha \beta}\,\Lam{k+l-N+i}_\gamma . \label{3.38}
\end{equation}
To get the last expression we also made use of the properties of the binomial
coefficients \cite{GrR}.

Note that
\begin{equation}
\X{k}{l}_{\alpha \beta} \equiv 0 \qquad {\rm for} \qquad k + l < N - 1 .
\label{3.39}
\end{equation}

Consider now the case of $N = 2$, which is distinguished from the
others by the fact that the structure functions depend on the
generalized coordinates $q^r$.
Performing the corresponding calculations, we see that this dependence
gives rise to the additional term to the expression (\ref{3.38}) for
$\X{k}{l}_{\alpha \beta}$.
This term is equal to $\dot q^r {\partial}/{\partial q^r}
\left(\A{N-l}{2N-k-l}^\gamma_{\alpha \beta}\right)\Lam{1}_\gamma$.

Thus, we have the following expression
for the Poisson brackets of the $k$-th and $l$-ary stage
secondary constraints $(k,l = 1,\ldots,N > 1)$
\begin{eqnarray}
\{ \Ph{k}_\alpha\,,\,\Ph{l}_\beta \} &\doteq& \left( \u{k}^\delta_\alpha\,
\A{1}{N-l+1}^\gamma_{\beta \delta} - \u{l}^\delta_\beta\,
\A{1}{N-k+1}^\gamma_{\alpha \delta} + \dot q^r \frac{\partial}{\partial q^r}\,
\A{N-l}{2N-k-l}^\gamma_{\alpha \beta} \right) \Lam{1}_\gamma \nonumber \\
&+& \sum_{i=0}^2 \sum_{j=0,1} {\left( \begin{array}{c}
                                     2N-k-l-i \\ N-l-j
                                    \end{array} \right)}
\A{j}{i}^\gamma_{\alpha \beta}\,\Lam{k+l-N+i}_\gamma . \label{3.40}
\end{eqnarray}

Remember that we have obtained all the expressions for the Poisson brackets
only on the primary constraint surface. Hence, these formulae determine the
relations of the constraint algebra up to linear combinations of the
primary constraints. To have the constraint algebra in the total phase space,
it is necessary to define a way of extension of functions from the
primary constraint surface to the whole phase space. One of such ways,
called the standard extension \cite{PR}, will be described in the next
Section.

Summarizing the above consideration, we see that the gauge invariance of the
form (\ref{2.4}), (\ref{2.5}) gives rise to the singular system with the set
of $N \times A$ projectable Lagrangian constraints.
Note that we have also proved that there appear in Hamiltonian
description of such systems $(N+1) \times A$ constraints being in
involution, at least on the primary constraint surface.

\section{The Poisson brackets within the standard extension}

Recall now main results and definitions dealing with the notion of the
standard extension. Following Ref.\cite{PR}, introduce the set of functions
$\chi^\alpha(q,\dot q)$, $\alpha = 1,\ldots,A$, such that
\begin{equation}
\u{0}^\delta_\alpha \ps{0}^r_\delta
\frac{\partial \chi^\beta}{\partial \dot q^r} = \delta^\beta_\alpha ,
\label{4.1}
\end{equation}
i.e. the vectors ${\partial \chi^\alpha}/{\partial \dot q^r}$ are dual to the
vectors $\u{0}^\delta_\alpha \ps{0}^r_\delta$.

The quite nontrivial aspects of existence of the functions $\chi^\alpha$,
satisfying (\ref{4.1}) were discussed in Ref.\cite{PR} (see also
\cite{NPR,NR2,NR3}) for the case of $N = 1$. Following to that discussion,
we assume the conditions of the existence of the functions
$\chi^\alpha$ to be valid.

Choose the functions $\chi^\alpha(q,\dot q)$ as follows:
\begin{equation}
\chi^\alpha(q,\dot q) = \dot q^r \chi^\alpha_r(q) + \nu^\alpha(q) .
\label{4.2}
\end{equation}
Hence, the vectors $\chi^\alpha_r(q)$ have to satisfy the duality relations
\begin{equation}
\u{0}^\delta_\alpha \ps{0}^r_\delta \chi^\beta_r = \delta^\beta_\alpha ,
\label{4.3}
\end{equation}
whereas $\nu^\alpha$ are some (arbitrary) functions of the generalized
coordinates $q^r$. We call a function $F(q,p)$ standard if
\cite{PR,NPR,NR1,NR2,NR3}
\begin{equation}
\chi^\alpha_r \frac{\partial F}{\partial p_r} = 0 . \label{4.4}
\end{equation}

One can show that for any function, defined on the primary constraint
surface, there exists a unique extension to the total phase space, which
is a standard function. This extension is called the standard extension.
The standard function coinciding with a function $F(q,p)$ on the
primary constraint surface is denoted by $F^0$. It is clear that for
any projectable function $f(q,\dot q)$ one can find an unique standard
function $F(q,p)$, such that $F \doteq f$. We denote this standard
function by $f^0$.

Using the properties of the standard extension \cite{PR}, one can obtain the
expression for the Poisson brackets of two standard functions $F^0$ and
$G^0$ :
\begin{equation}
\{ F^0\,,\,G^0 \} = \{ F\,,G \}^0 + \frac{\partial F^0}{\partial p_r}
\chi^\alpha_{rs} \frac{\partial G^0}{\partial p_s} \, \Ph{0}_\alpha ,
\label{4.5}
\end{equation}
where
\begin{equation}
\chi^\alpha_{rs} = \frac{\partial \chi^\alpha_r}{\partial q^s} -
\frac{\partial \chi^\alpha_s}{\partial q^r} . \label{4.6}
\end{equation}
Note that we have introduced  the so-called standard primary
constraints in Eq.(\ref{4.5}), defined by the relations
\begin{equation}
\Ph{0}_\alpha \doteq 0 , \qquad \frac{\partial \Ph{0}_\alpha}{\partial p_r}
= - ( \u{0}^\beta_\alpha \ps{0}^r_\beta )^0 . \label{4.7}
\end{equation}

Let the Hamiltonian $H$ and all the constraints of the system $\Ph{0}_\alpha$,
$\Ph{k}_\alpha$, $k = 1,\ldots,N$, be the standard functions. Then, using
Eqs.(\ref{3.27})--(\ref{3.30}), (\ref{3.40}), we get from (\ref{4.5}) that
the constraint algebra of the system under consideration is given by the
expressions
\begin{eqnarray}
&&\{ H\,,\,\Ph{0}_\alpha \} \;=\, \bigl(\u{0}^\beta_\alpha\bigr)^0\;
\Ph{1}_\beta +
\frac{\partial H}{\partial p_r}\chi^\beta_{rs}
\frac{\partial \Ph{0}_\alpha}{\partial p_s}\;\Ph{0}_\beta ,
\label{4.8} \\
&&\{ H\,,\,\Ph{k}_\alpha \} \;=\, \Ph{k+1}_\alpha -
\Bigl( \u{k}^\beta_\alpha + \mu^\delta \A{1}{N-k+1}^\beta_{\alpha \delta}
\Bigr)^0\; \Ph{1}_\beta + \frac{\partial H}{\partial p_r}\chi^\beta_{rs}
\frac{\partial \Ph{k}_\alpha}{\partial p_s}\; \Ph{0}_\beta ,
\label{4.9} \\
&&\{ \Ph{0}_\alpha\,,\,\Ph{0}_\beta \} \,=\,
\frac{\partial \Ph{0}_\alpha}{\partial p_r}\chi^\gamma_{rs}
\frac{\partial \Ph{0}_\beta}{\partial p_s}\;\Ph{0}_\gamma ,
\label{4.10} \\
&&\{ \Ph{k}_\alpha\,,\,\Ph{0}_\beta \} \,=\, \Bigl( \u{0}^\delta_\beta\,
\A{1}{N-k+1}^\gamma_{\alpha \delta} \Bigr)^0\;\Ph{1}_\gamma +
\frac{\partial \Ph{k}_\alpha}{\partial p_r}\chi^\gamma_{rs}
\frac{\partial \Ph{0}_\beta}{\partial p_s}\;\Ph{0}_\gamma ,
\label{4.11} \\
&&\{ \Ph{k}_\alpha\,,\,\Ph{l}_\beta \} \,=\, \Bigl( \u{k}^\delta_\alpha\,
\A{1}{N-l+1}^\gamma_{\beta \delta} - \u{l}^\delta_\beta\,
\A{1}{N-k+1}^\gamma_{\alpha \delta} + \dot q^r \frac{\partial}{\partial q^r}\,
\A{N-l}{2N-k-l}^\gamma_{\alpha \beta} \Bigr)^0\; \Ph{1}_\gamma \nonumber \\
&+& \sum_{i=0}^2 \sum_{j=0,1} {\left( \begin{array}{c}
                                     2N-k-l-i \\ N-l-j
                                    \end{array} \right)}
\A{j}{i}^\gamma_{\alpha \beta}\;\Ph{k+l-N+i}_\gamma
+ \frac{\partial \Ph{k}_\alpha}{\partial p_r}\chi^\gamma_{rs}
\frac{\partial \Ph{l}_\beta}{\partial p_s}\;\Ph{0}_\gamma ,
\label{4.12}
\end{eqnarray}
for any $k, l = 1,\ldots,N > 1$, and from (\ref{3.39}) we have  that
$i > N - k - l$.
Besides, the functions $\mu^\alpha(q,\dot q)$ and $\u{k}(q,\dot q)$
are of the form
\begin{eqnarray}
\mu^\alpha &=& \dot q^r \chi_r^\beta \u{0}^\alpha_\beta ,
\label{4.13} \\
\u{k}^\alpha_\beta &=& \ps{N-k}^r_\beta \chi_r^\gamma
\u{0}^\alpha_\gamma .  \label{4.14}
\end{eqnarray}
It is seen from Eq.(\ref{4.10}), that the primary constraints form a
subalgebra of the constraint algebra.

Now let us compute the Poisson brackets of the standard Hamiltonian and
constraints with arbitrary standard function $F$. It is useful to have the
corresponding formulas from the point of view of possible applications
(see e.g. \cite{NR2,NR3,Hen}). To obtain these expressions, define
the projector
\begin{equation}
\Pi^r_s = \delta^r_s - \chi^\alpha_s \u{0}^\beta_\alpha \ps{0}^r_\beta ,
 \qquad \Pi^t_s \Pi^r_t = \Pi^r_s , \label{4.15}
\end{equation}
and introduce the so-called pseudo-inverse matrix $W^{rs}(q,\dot q)$
\cite{L,R} for the singular Hessian $W_{rs}(q,\dot q)$. It can be shown
\cite{R} that for any singular matrix $W_{rs}$ there exists a
pseudo-inverse matrix $W^{rs}$, defined uniquely by the relations
\begin{eqnarray}
W^{rt} W_{ts} &=& \Pi^r_s , \label{4.16} \\
W^{rs} \chi_s^\alpha &=& 0 . \label{4.17}
\end{eqnarray}
Consider a standard function $F(q,p)$ on the phase space . It is clear that
there exists a function $f(q,\dot q)$ on the velocity phase space,
connected with $F$ by the relation (\ref{3.9}). Using the definitions of a
standard function and the pseudo-inverse matrix (\ref{4.4}),
(\ref{4.15})--(\ref{4.17}),
we obtain the following expressions for the partial derivatives of the
standard
function $F$ \cite{NPR,NR3}
\begin{eqnarray}
\frac{\partial F}{\partial p_r} &\doteq& W^{rs}
\frac{\partial f}{\partial \dot q^s} , \label{4.18} \\
\frac{\partial F}{\partial q^r} &\doteq& \frac{\partial f}{\partial q^r}
- \frac{\partial f}{\partial \dot q^s} W^{st}
\frac{\partial^2 L}{\partial \dot q^t \partial q^r} , \label{4.19}
\end{eqnarray}
Now, making use of the expressions for the partial derivatives of the
standard functions $F$, $H$, $\Ph{0}_\alpha$, $\Ph{k}_\alpha$,
$k = 1,\ldots,N > 1$, we get the Poisson brackets of the form
\begin{eqnarray}
\{ H\,,\,F \} &\doteq& -\, T(f) + \mu^\alpha\left( \ps{0}^r_\alpha
\frac{\partial f}{\partial q^r} + \Bigl(\ps{1}^r_\alpha +
T(\ps{0}^r_\alpha)\Bigr)
\frac{\partial f}{\partial \dot q^r} \right) \nonumber \\
&& -\, \mu^\alpha v^\beta_\alpha \Bigl(\frac{\partial f}{\partial \dot q^r}
W^{rs} \frac{\partial \u{0}^\gamma_\beta}{\partial \dot q^s}\Bigr)\,
\Lam{1}_\gamma ,
\label{4.20} \\
\{ \Ph{0}_\alpha\,,\,F \} &\doteq&  \u{0}^\beta_\alpha\left( \ps{0}^r_\beta
\frac{\partial f}{\partial q^r} + \Bigl(\ps{1}^r_\beta +
T(\ps{0}^r_\beta)\Bigr)
\frac{\partial f}{\partial \dot q^r} \right)
 -\, \Bigl(\frac{\partial f}{\partial \dot q^r}
W^{rs} \frac{\partial \u{0}^\beta_\alpha}{\partial \dot q^s}\Bigr)\,
\Lam{1}_\beta ,
\label{4.21}
\end{eqnarray}
where $v^\beta_\alpha$ is the inverse matrix for the
matrix $\u{0}^\alpha_\beta$, and
\begin{eqnarray}
\{ \Ph{k}_\alpha\,,\,F \} &\doteq&  \ps{k}^r_\alpha
\frac{\partial f}{\partial q^r} + \Bigl(\ps{k+1}^r_\alpha +
T(\ps{k}^r_\alpha)\Bigr)
\frac{\partial f}{\partial \dot q^r}
 \nonumber \\
&-& \u{k}^\beta_\alpha \left( \ps{0}^r_\beta
\frac{\partial f}{\partial q^r}  + \Bigl( \ps{1}^r_\beta +
T(\ps{0}^r_\beta) \Bigr) \frac{\partial f}{\partial \dot q^r} \right)
+ \Bigl(\frac{\partial f}{\partial \dot q^r} W^{rs} \frac{\partial
\u{k}^\beta_\alpha}{\partial \dot q^s}\Bigr)\, \Lam{1}_\beta  .
\label{4.22}
\end{eqnarray}
Here we have introduced the notation
\begin{equation}
T = \dot q^t\frac{\partial}{\partial q^t} +
R_s W^{st}\frac{\partial}{\partial \dot q^t} . \label{4.23}
\end{equation}
Note that on the Lagrange equations $L_r = 0$ we have
\begin{equation}
T(f) = \frac{d}{dt}(f), \label{4.24}
\end{equation}
so the differential operator $T$ is nothing but the evolution operator
of gauge invariant systems.

Note again that the Eqs.(\ref{4.20})--(\ref{4.22}) determine the Poisson
brackets only on the primary constraint surface. To obtain the
corresponding formulae in the total phase space, it is sufficient to
apply Eq.(\ref{4.5}) for standardly extended Poisson brackets to these
expressions.

\newpage

\section{Conclusion}

In this paper we have considered the mechanical systems, which are invariant
under gauge transformations of the form (\ref{2.4}) and established the
correspondence between Lagrangian and Hamiltonian descriptions of such
systems. On the base of the notion of the standard extension we have obtained
the explicit form of the constraint algebra, the latter turned out to
be the first class.

The results of this paper and Refs.\cite{PR,NPR} complete the
consistent analysis of gauge invariant systems of general form, where
only projectable Lagrangian constraints appear.

Note finally that the gauge transformations (\ref{2.4}) are mapped to the
phase space as follows (cf. Refs.\cite{M,PR,HTZ,P})
\begin{eqnarray}
\delta q^r &\doteq& \{ q^r\,,\,G_\varepsilon \} , \label{5.1} \\
\delta p_r &\doteq& \{ p_r\,,\,G_\varepsilon \} +
\frac{\partial \delta q^s}{\partial \dot q^r} \,L_s , \label{5.2}
\end{eqnarray}
Where $G_\varepsilon$ is the linear combination of the constraints
\begin{eqnarray}
G_\varepsilon &=& \sum_{k=0}^N {\g{k}^\alpha(\varepsilon)\,\Ph{k}_\alpha} ,
\label{5.3} \\
\g{0}^\alpha(\varepsilon) &=& - \left(\eps{N}^\beta +
\sum_{k=0}^{N-1} {\eps{k}^\gamma\, \u{N-k}^\beta_\gamma} \right)
v^\alpha_\beta ,
\label{5.4} \\
\g{k}^\alpha(\varepsilon) &=& -\, \eps{N-k} , \qquad k = 1,\ldots,N .
\label{5.5}
\end{eqnarray}

\vskip1cm

The author is grateful to Profs. A.V. Razumov, V.A. Rubakov and F.V. Tkachov
for support, useful discussions and valuable remarks. He is also indebted to
A.N. Kuznetsov, G.B. Pivovarov and A.V. Subbotin for helpful
discussions of the results.  This research was supported in part by
the International Science Foundation under grant  MP 9000 .

\end{document}